\documentclass[conference,twocolumn]{IEEEtran}
\IEEEoverridecommandlockouts

\def\paperversion{2} % 1 - conference version, 2 - full version

\usepackage{etex}
\usepackage[subtle,indent=normal,mathspacing=normal,wordspacing=normal,title=normal,charwidths=normal]{savetrees}

\usepackage{balance}

\usepackage[dvipsnames]{xcolor}

\usepackage{multirow}
\usepackage[cmex10]{amsmath}
\usepackage{amssymb,amsfonts,amssymb,amsthm}
\usepackage{stmaryrd}
\usepackage{mathtools}
\usepackage{bbm}
\usepackage[T1]{fontenc}
\usepackage{cite}
\usepackage{algorithmic}
\usepackage{graphicx}
\usepackage{textcomp}
\usepackage{algorithm2e}
\RestyleAlgo{ruled}
\SetKwComment{Comment}{/* }{ */}
\SetKwInput{KwInput}{Input}
\SetKwInput{KwOutput}{Output}

\usepackage{url}
\usepackage{enumerate}
\usepackage{float}
\usepackage{booktabs}
\usepackage{multirow}
\usepackage{colortbl}
\usepackage[font=footnotesize]{caption}
\usepackage[font=footnotesize]{subcaption}

\usepackage{cleveref}
\Crefname{figure}{Fig.}{Figs.}

\usepackage{pgfplots}
\usepackage{pgfplotstable}
\usepackage{tikz}
\usetikzlibrary{calc,decorations.pathreplacing}
\usepackage{cleveref}
\DeclareCaptionLabelSeparator{periodspace}{.\quad}
\def\BibTeX{{\rm B\kern-.05em{\sc i\kern-.025em b}\kern-.08em
    T\kern-.1667em\lower.7ex\hbox{E}\kern-.125emX}}

\newtheorem{theorem}{Theorem}
\newtheorem{proposition}{Proposition}

\makeatletter
\newcommand{\removelatexerror}{\let\@latex@error\@gobble}
\makeatother

\usepackage[left=1.62cm,right=1.62cm,top=0.75in,bottom=1.05in]{geometry}

\begin{document}

\title{Sparse Degree Optimization for BATS Codes}
\author{%
  \IEEEauthorblockN{Hoover~H.~F.~Yin and Jie Wang}
  \thanks{H.~H.~F.~Yin is with the Department of Information Engineering, The Chinese University of Hong Kong.
  J.~Wang is with the H. Milton Stewart School of Industrial and Systems Engineering, Georgia Institute of Technology.
  }
}

\maketitle

\begin{abstract}
Batched sparse (BATS) code is a class of batched network code that can achieve a close-to-optimal rate when an optimal degree distribution is provided.
We observed that most probability masses in this optimal distribution are very small, i.e., the distribution ``looks'' sparse.
In this paper, we investigate the sparsity optimization of degree distribution for BATS codes that produces sparse degree distributions.
There are many advantages to use a sparse degree distribution, say, it is
robust to precision errors when sampling the degree distribution during encoding and decoding in practice.
We discuss a few heuristics and also a way to obtain an exact sparsity solution.
These approaches give a trade-off between computational time and achievable rate, thus give us the flexibility to adopt BATS codes in various scenarios, e.g., device with limited computational power, stable channel condition, etc.
\end{abstract}

\section{Introduction}

In a traditional network communication scenario, packet loss is mainly due to congestion.
In a wireless network, there are more ways to lose a packet, e.g., interference, fading signals, etc.
To achieve reliable transmission, it is a common practice to apply end-to-end retransmission and congestion control strategies.
However, these strategies are not suitable for multihop wireless networks with packet loss, as a packet can arrive at the destination if it is not being dropped at any of the links.
This means that an end-to-end retransmission scheme needs to send a huge amount of packets, thus the capacity of the network cannot be achieved.

For a huge range of scenarios, the capacity of a network 
can be achieved by a realization of network coding \cite{linear,flow} known as \emph{random linear network coding (RLNC)} \cite{random2}.
Instead of applying store-and-forward strategy at the intermediate network nodes, recoding (re-encoding) is performed.
Simply speaking, every packet transmitted by an intermediate network node is a random linear combination of the packets received by the node.
This way, the node can send more packets than what it has received.
However, a direct implementation of RLNC has a few practical concerns, including high storage and computational costs at the intermediate network nodes, and a huge coefficient vector overhead when the data is long.

\emph{Batched sparse (BATS) codes} \cite{yang14bats,bats_book} is a variation of RLNC that aims to resolve the these issues.
A BATS code consists of an outer code and an inner code.
The outer code is a matrix generalization of fountain codes \cite{lubyLT} that encodes the data into \emph{batches}, where each batch contains a few coded packets.
The inner code restricts the application of RLNC within the packets belonging to the same batch.
Similar as fountain codes, the encoding of batches depends on a \emph{degree distribution}.
When the degree distribution is optimized, BATS codes have a close-to-optimal rate.
Unlike fountain codes, there is no universal degree distribution when each batch consists of more than one coded packets \cite{quasi}.
In other words, we need to optimize the degree distribution to achieve the best rate of BATS codes.
This is known as the \emph{degree optimization problem}.

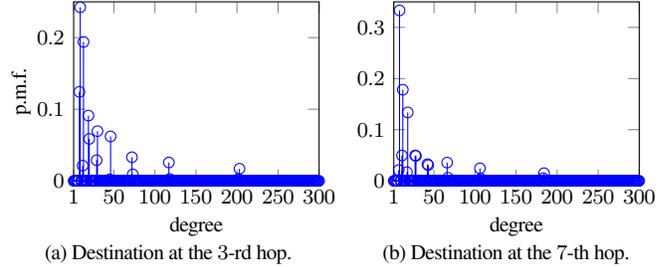
\begin{figure}
    \centering
    \hskip -1.1em
    \begin{subfigure}{.24\textwidth}
    \centering
    \begin{tikzpicture}
		\begin{axis}[
			small,
                %scale only axis,
                width=1.1\textwidth,
                height=.9\textwidth,
			xtick={1,50,100,150,200,...,700},
			ytick={0.0,0.1,...,0.3},
			xmin=1, xmax=300, %799,
			ymin=0.0, ymax=0.25,
			xlabel=degree,
			ylabel=p.m.f., %probability mass function,
			x label style={at={(axis description cs:.5,.07)},anchor=north},
			y label style={at={(axis description cs:.24,.5)},anchor=south},
                label style={font=\footnotesize}
		]
			\addplot+[ycomb,mark=o] table [x=D, y=n3, col sep=comma] {degdist.csv};
		\end{axis}
    \end{tikzpicture}
    \vskip -.7em
    \caption{Destination at the $3$-rd hop.}
    \label{fig:deg1}
    \end{subfigure}~%\\[.5em]
    \begin{subfigure}{.24\textwidth}
    \centering
    \begin{tikzpicture}
		\begin{axis}[
			small,
                %scale only axis,
                width=1.1\textwidth,
                height=.9\textwidth,
			xtick={1,50,100,150,200,...,700},
			ytick={0.0,0.1,...,0.4},
			xmin=1, xmax=300, %799,
			ymin=0.0, ymax=0.35,
			xlabel=degree,
			%ylabel=p.m.f., %probability mass function,
			x label style={at={(axis description cs:.5,.07)},anchor=north},
			y label style={at={(axis description cs:.24,.5)},anchor=south},
                label style={font=\footnotesize}
		]
			\addplot+[ycomb,mark=o] table [x=D, y=n7, col sep=comma] {degdist.csv};
		\end{axis}
    \end{tikzpicture}
    \vskip -.7em
    \caption{Destination at the $7$-th hop.}
    \label{fig:deg2}
    \end{subfigure}
    \vskip -.2em
    \caption{Illustrations of optimal degree distributions.}%, which are sparse-looking.}
    %\vspace{-2.2em}
\end{figure}

This paper is motivated by our observation that the optimal degree distributions ``look'' sparse.
To illustrate our observation, we consider the following simple examples.
Suppose each batch has $8$ coded packets, and a BATS code is applied to transmit a piece of data through a few network links with $10\%$ independent packet loss rate each.
\Cref{fig:deg1} is the stem plot of the optimal degree distribution if the $3$-rd hop is the destination, and \Cref{fig:deg2} is the one if the $7$-th hop is the destination.
Besides the few distinctive probability masses, the remaining masses are close to $0$, yet not equal to $0$.
It is natural to ask: Can we construct a ``real'' sparse degree distribution that has a close-to-optimal achievable rate?

There are many advantages to adopting sparse degree distributions.
First, the close-to-$0$ masses are vulnerable to numerical or precision errors when sampling from the cumulative degree distribution.
In contrast, sampling from a sparse degree distribution is more numerically stable and efficient.
Second, as the support size of the sparse distribution is small, it provides us with an interpretable result on which degrees are crucial to achieve high rates.
Third, the encoder and decoder have to agree on the same degree distribution for correctness.
We need to either send the ingredients for the degree optimization problem or the optimized degree distribution to the other end of the network.
As this transmission is not protected by BATS codes, 
we want to send a short message to cope with the risk of packet loss, which is particularly important when one needs to frequently find a new degree distribution as the channel conditions shift over time.
This challenge can be tackled by considering a sparse degree distribution.

In this paper, we discuss a few heuristics to produce a close-to-optimal sparse degree distribution, and an exact solution via solving a mixed-integer program.
These approaches give a trade-off between computational time and achievable rate, thus give us the flexibility to adopt to various applications.
For example, if the channel condition and the recoding policy remain unchanged throughout the transmission of the data, then we can spend more time to compute a better degree distribution.
Otherwise, we may need to update the degree distribution to adopt to the situation, so we prefer a degree distribution that can be computed quickly.
We also record the computational time for each approach as a reference.
One surprising result is that one of our approaches (via complementary slackness) runs faster than solving the original non-sparse problem.
This gives an efficient alternative even when sparsity is not in concern.

As a remark, our approaches can be easily modified for other linear programming based degree optimization formulations for different applications, such as multiple receivers \cite{quasi}, expanding window \cite{expanding}, sliding window \cite{sliding}, unequal protection \cite{unequal}, distributed fog computing~\cite{yue2018distributed}, etc.
These formulations originally did not consider the sparsity and gave sparse-looking solutions.

\textbf{Notations:} 
Denote $\llbracket N \rrbracket := \{1, 2, \ldots, N\}$ for any positive integer $N$.
Denote by $I_x(a, b)$ %= \frac{\int_0^x t^{a-1} (1-t)^{b-1} dt}{\int_0^1 t^{a-1} (1-t)^{b-1} dt}$ 
the regularized incomplete beta function. %, which can be used to express partial sum of probability masses of a binomial distribution.
%We have
%\begin{equation*}
%    I_x(d-r, r) = \sum_{j = d-r}^{d-1} \binom{d-1}{j} x^j (1-x)^{d-j-1}.
%\end{equation*}
Let $\mathbb{E}[\cdot]$ be the expectation operator.
The $m \times 1$ zero vector and all-one vector are denoted by $\mathbf{0}_m$ and $\mathbf{1}_m$ respectively.
Given a vector $\mathbf{x} = (x_1, x_2, \ldots, x_m)^T \in \mathbb{R}^m$, denote by $\|\mathbf{x}\|_0 = \sum_{i = 1}^m \mathbbm{1}_{x_i \neq 0}$ the $\ell_0$-norm of $\mathbf{x}$, where $\mathbbm{1}$ is the indicator function.
Fix a finite field $\mathbb{F}_q$ of size $q$, and fix a positive integer $M$ called the \emph{batch size}.

\section{BATS Codes}

\subsection{Encoding}

Before encoding the data to be transmitted by a BATS code, we can first apply a \emph{precode} to the data.
A precode is simply an erasure code.
By doing so, the BATS code only needs to recover a sufficient portion of the precoded data, and the precode can then recover the original data.
This technique, proposed for Raptor codes \cite{shokRaptor}, is useful for codes that rely on belief propagation (BP).
It can maintain a constant decoding complexity with respect to the data size after belief propagation (BP) decoding has been stopped.

Now, we apply a BATS code to the precoded data.
The data is first divided into multiple \emph{input packets}, where each input packet is regarded as a vector over $\mathbb{F}_q$ of the same length.
The length of the input packets can be optimized via approaches in \cite{pktsize} to minimize the padding overhead in practice.
A BATS encoder generates a sequence of batches, where each batch consists of $M$ coded packets.

Each batch is generated as follows.
First, a \emph{degree} is sampled from a predefined \emph{degree distribution}, where the value of the degree, $d$, is the number of input packets to be chosen to form the batch.
The degree distribution must be optimized in order to achieve the best rate.
We defer the discussion of the degree optimization problem to \Cref{sec:degree}.
After obtaining the degree $d$ for the batch, the encoder selects $d$ input packets uniform randomly.
Each of the $M$ packets in the batch is a random linear combination of the chosen $d$ input packets.

These $M$ packets are defined to be linearly independent of each other by the mean of coefficient vectors.
Two packets are linearly independent of each other if and only if their coefficient vectors are linearly independent of each other.
That is, a coefficient vector is attached to each packet in the batch, and it is initialized in a way to ensure the linear independence \cite{yang22pro,yin20pro}.
The number of linearly independent packets in a batch is called the \emph{rank} of the batch, which is a measure for the information remained in the batch.

To reduce the transmission overhead, a common practice is to use the batch ID as a seed for a pseudorandom number generator to reproduce the randomness used during encoding, i.e., the degree and the random coefficients for linear combinations.
This way, both the encoder and decoder must agree with the same degree distribution before the channel is protected by BATS codes.

\subsection{Recoding and Decoding}

At each intermediate network node, recoding is applied to a batch after the packets of this batch that are not lost are received.
The recoding operations are restricted to the packets belonging to the same batch.
During recoding, recoded packets are generated.
Each recoded packet of a batch is a random linear combination of the received packets of the batch.
This time, the coefficient vector and the coded data of the packet are concatenated and regarded as a single vector for random linear combinations, so that the recoding operations can be recorded.

After generating certain number of recoded packets for a batch, they are transmitted to the next node.
The batch can then be discarded by the current node from its buffer.
The number of recoded packets to be generated depends on the recoding scheme.
The simplest scheme is to generate the same number of recoded packets for each batch.
This is known as \emph{baseline recoding}, which is applied in many works such as \cite{variable,fun,delay,buffer,zhou17b,bats_schedule} due to its simplicity.
However, this is not an optimal scheme \cite{yang14a}.
\emph{Adaptive recoding} \cite{adaptive,scheduling,ge_adaptive,uni,bar} decides the number of recoded packets according to the rank of the batch so that the expected rank of the batches at the next node, i.e., the expected amount of information carried by the batches, is maximized.

At the destination node, the received batches are passed to a belief propagation (BP) decoder to recover the input packets.
If the rank of the batch is no smaller than the degree of the batch, we can recover the involved input packets by solving a system of linear equations.
The decoded input packets are then substituted to other received batches that have not decoded yet.
The effective degrees of these batches are decreased.
If any of these batches can be decoded, then the recovered input packets will be substituted to other batches, and so on and so forth.
Inactivation decoding \cite{shokRaptor,Raptormono} can also be used to continue the decoding procedure after BP decoding has been stopped.
The whole decoding procedure stops once a sufficient amount of input packets are recovered for the precode to recover the original data.

The ranks of the batches arriving at the destination node forms a \emph{rank distribution}.
The expected value of this rank distribution is the theoretical upper bound on the achievable rates \cite{yang11x2}, where a BATS code with an optimized degree distribution can achieve a rate very close to this value.
Therefore, this rank distribution is an ingredient for the degree optimization problem.

\subsection{Degree Distribution Optimizations}
\label{sec:degree}

Now, we present the basic formulation for optimizing the degree distribution.
We desire a degree distribution $(\Psi_1, \Psi_2, \ldots, \Psi_D)$ that can maximize the achievable rate $\theta$ of a BATS code.
The positive integer $D$ is the maximum degree.
The rate will not improve for $D > \lceil M/(1-\eta) \rceil  - 1$ \cite[Thm.~6.2]{bats_book}, where $\eta \in (0, 1)$ is the sufficient portion of the precoded data for the precode to recover the original data.
Therefore, we set $D = \lceil M/(1-\eta) \rceil - 1$ in the remaining text.
Define
\begin{equation*}
\small{
    \mathcal{P} = \left\{ 
     (\Psi_1, \Psi_2, \ldots, \Psi_D)^T \colon \sum_{d = 1}^D \Psi_d = 1, \Psi_d \ge 0, \forall d \in \llbracket D \rrbracket 
    \right\}}.
\end{equation*}

To maximize $\theta$, we need the knowledge of the rank distribution $(h_0, h_1, \ldots, h_M)$ at the destination node.
Define $ \hbar_k = \sum_{i = k}^M \frac{\zeta_k^i}{q^{i-k}} h_i,$
which is the probability that a batch is decodable for the first time when its degree is $k$, where
\begin{equation*}
    \zeta_r^m = \begin{cases}
        \prod_{i = 0}^{r-1} (1-q^{-m+i}) & \text{if } 0 < r \le M,\\
        1 & \text{otherwise}
    \end{cases}
\end{equation*}
is the probability of an $r \times m$ totally random matrix over $\mathbb{F}_q$ is full rank \cite{gadouleau10packing}.
When the field size is large enough, e.g., $q = 2^8$, the difference between $h_k$ and $\hbar_k$ becomes negligible.
For brevity, we omit $h_0$ and $\hbar_0$ and write $\pmb{\hbar} = (\hbar_1, \hbar_2, \ldots, \hbar_M)^T$.

For each $x \in (0, \eta]$ where $\eta < 1$, define an $M \times D$ matrix $\mho(x)$ where its $(r,d)$-th entry is
\begin{equation*}
    (\mho(x))_{r,d} = \begin{cases}
        d & \text{if } d \le r,\\
        d I_x(d-r, r) 
        & \text{if } d > r.
    \end{cases}
\end{equation*}
The degree optimization problem for a single rank distribution $\mathbf{h}$ is a linear programming problem
\begin{equation}
    \smashoperator{\max_{\boldsymbol{\Psi} \in \mathcal{P}, \theta \in \mathbb{R}}} \,\,\,\,\theta \quad \mathrm{s.t.} \quad \inf_{x \in (0, \eta]} (\pmb{\hbar}^T \mho(x) \boldsymbol{\Psi} + \theta \ln(1-x)) \ge 0,
    \label{Eq:optimization:nominal}
\end{equation}
where the constraint gives the necessary and sufficient condition for decoding up to $\eta$ portion of the data.
This is a known formulation obtained via differential equation analysis \cite{yang14bats} or tree analysis \cite{tree}.
Problem~\eqref{Eq:optimization:nominal} essentially contains uncountably infinite number of constraints, which is called \emph{semi-infinite} optimization in literature.
Existing works \cite{yang14bats,bats_book,quasi,sliding,expanding,unequal,variable} consider $x$ in a set of discrete parameters drawn from the interval $(0, \eta]$ with a sufficiently small step size, denoted as $\mathcal{X}$, and solve the following approximation problem instead:
\begin{equation}
    \smashoperator{\max_{\boldsymbol{\Psi} \in \mathcal{P}, \theta \in \mathbb{R}}} \,\,\,\,\theta \quad \mathrm{s.t.} \quad \min_{x \in \mathcal{X}} (\pmb{\hbar}^T \mho(x) \boldsymbol{\Psi} + \theta \ln(1-x)) \ge 0,
    \label{Eq:optimization:nominal:app}
\end{equation}
The step size affects the optimal rate $\theta$ and the number of constraints in the linear program.
The following proposition and theorem quantifies the convergence behavior of the optimal solution and optimal value between Problem~\eqref{Eq:optimization:nominal} and  its discrete approximation by taking $x\in\mathcal{X}$.
To begin with, we specify $\mathcal{X}=\{x_i\}_{i=1}^m$ that consists of $m$ parameters, and the mesh size $\rho_m$ that quantifies the tightness of discretization of the interval $(0,\eta]$, i.e.,
$
\rho_m = \sup_{x\in(0,\eta]}\min_{i \in \llbracket m \rrbracket} |x-x_i|.
$
Denote by $\theta_m$ the $\epsilon_m$-optimal solution to \eqref{Eq:optimization:nominal:app}, i.e., $\theta_m$ is feasible in \eqref{Eq:optimization:nominal:app} and $\theta_m\ge \text{Optval}\eqref{Eq:optimization:nominal:app} - \epsilon_m$.

\begin{proposition}[Convergence of Discretization]\label{Proposition:converge}
Assume that as $m\to\infty$, $\epsilon_m\downarrow 0$ and $\rho_m\downarrow0$.
Then any accumulation point of the sequence $\{\theta_m\}_m$ is an optimal solution to \eqref{Eq:optimization:nominal}.
\end{proposition}

\ifnum\paperversion=2
\begin{IEEEproof}
    See Appendix~\ref{Sec:proof:converge}.
\end{IEEEproof}
\fi

\begin{theorem}[Convergence Rate of Discretization]\label{Theorem:converge:rate}
Under the same assumption as in Proposition~\ref{Proposition:converge}, assume in addition that the optimal solution to \eqref{Eq:optimization:nominal} is a singleton, denoted as $\{(\theta^*, \boldsymbol{\Psi}^*)\}$, and $\epsilon_m=\mathcal{O}(\rho_m)$, then it holds that 
$|\theta_m - \theta^*| = \mathcal{O}(\rho_m)$.
\end{theorem}

\ifnum\paperversion=2
\begin{IEEEproof}
    See Appendix~\ref{Sec:proof:converge}.
\end{IEEEproof}
\fi

\section{Sparsity Optimizations}\label{Sec:sparse:opt}

Although we observed that the optimal degree distribution looks sparse, the small masses still have an effect to the optimal rate.
In other words, when we optimize for the sparsity, we expect a small rate drop.
We consider a few heuristics to obtain a sparse degree distribution, which impose a trade-off between computational time and rate drop.
We also discuss an exact solver for the sparsity problem.
The numerical comparison is deferred to the next section.

Note that the computational cost is in consideration if the destination nodes re-solve the optimization frequently with updated empirical rank distributions, e.g., for time-varying channels that have non-constant packet loss rate, for varying recoding policy at intermediate nodes due to non-constant transmission rates, etc.

\subsection{Via Direct Trimming}\label{Sec:direct}

As the optimal $\boldsymbol{\Psi}$ given by \eqref{Eq:optimization:nominal} looks sparse already, the simplest approach is to trim the masses smaller than a threshold $\epsilon > 0$ to $0$ directly and then normalize the remaining non-zero masses.
This approach has the least effort among the approaches in this paper, but the sparsity and the rate are not good enough.

\subsection{Via Complementary Slackness}

We first consider the dual formation of \eqref{Eq:optimization:nominal} after discretizing the interval $(0, \eta]$ into a set $\mathcal{X}$:
\begin{equation*}\small{
    \begin{IEEEeqnarraybox}[][c]{rCl}
        \min_{\mu \in \mathbb{R}, \boldsymbol{\lambda} \ge \mathbf{0}_{|\mathcal{X}|}, \boldsymbol{\gamma} \ge \mathbf{0}_D} & \quad & \mu\\
        \mathrm{s.t.} && \sum_{x \in \mathcal{X}} \lambda_x \pmb{\hbar}^T \mho(x) - \mu \mathbf{1}_{D}^T + \boldsymbol{\gamma}^T = \mathbf{0}_{D}^T\\
        && 1 + \sum_{x \in \mathcal{X}} \lambda_x \ln(1-x) = 0.
    \end{IEEEeqnarraybox}}
\end{equation*}

Let $\theta^\ast, \boldsymbol{\Psi}^\ast$ be the optimal primal variables and $\mu^\ast, \boldsymbol{\lambda}^\ast, \boldsymbol{\gamma}^\ast$ be the optimal dual variables.
By complementary slackness of the KKT condition, we must have the Hadamard product $\boldsymbol{\gamma}^\ast \odot \boldsymbol{\Psi}^\ast = \mathbf{0}_D.$
Although we do not know the optimal $\boldsymbol{\Psi}^\ast$ by solving the dual problem solely, the optimal $\boldsymbol{\gamma}^\ast$ gives us some hints on the sparsity of $\boldsymbol{\Psi}^\ast$.
More specifically, if $\gamma^\ast_d > 0$, then we must have $\Psi^\ast_d = 0$.
Note that when $\gamma^\ast_d = 0$, we have no conclusion on the value of $\Psi^\ast_d$.

Due to numerical precision of the solver for solving the dual problem, we may set a threshold $\epsilon$ so that we consider $\Psi_d = 0$ if $\gamma^\ast_d \ge \epsilon$.
In other words, we filter the support of $\boldsymbol{\Psi}$ we are interested as the set $\mathcal{S} := \{ d \in \llbracket D \rrbracket \colon \gamma^\ast_d < \epsilon \}.$

Due to strong duality of linear programs, the optimal value $\mu^\ast$ equals to the optimal rate $\theta^\ast$ of the primal problem, thus the remaining problem is to find a feasible $\boldsymbol{\Psi}$ over the support $\mathcal{S}$ for the primal problem.
Define $\boldsymbol{\Psi}'_\mathcal{S} = \{\Psi'_d \colon d \in \mathcal{S}\}$.
Let $(\pmb{\hbar}^T \mho(x))_\mathcal{S}$ be a row vector by removing those columns of $\pmb{\hbar}^T \mho(x)$ that column indices are not in $\mathcal{S}$.
We first obtain a feasible $\boldsymbol{\Psi}'_\mathcal{S}$ by solving %the following linear program:
\begin{equation*}
\small{
    \begin{IEEEeqnarraybox}[][c]{rClcl}
        \min_{\boldsymbol{\Psi}'_\mathcal{S}} & \quad & 1\\
        \mathrm{s.t.} && (\pmb{\hbar}^T \mho(x))_\mathcal{S} \boldsymbol{\Psi}'_\mathcal{S} + \mu^\ast \ln(1-x) \ge 0, & \quad & \forall x \in \mathcal{X}\\
        && \sum_{d \in \mathcal{S}} \Psi'_d = 1, \quad  \Psi'_d \ge 0, && \forall d \in \mathcal{S}.
    \end{IEEEeqnarraybox}}
\end{equation*}
Then, we obtain a sparse $\boldsymbol{\Psi}$ by $\Psi_d = \Psi'_d$ if $d \in \mathcal{S}$, or $\Psi_d = 0$ otherwise.

\subsection{Iterative Reweighted $\ell_1$-Norm Heuristic}\label{Sec:ell:1}

Suppose we want to obtain a sparse $\boldsymbol{\Psi}$ that can (approximately) achieve a target rate $\theta$.
This rate must be a achievable one, i.e., no larger than the optimal rate obtained by solving \eqref{Eq:optimization:nominal}, leading to the optimization 
\[
\vspace{-0.65em}
\min_{\boldsymbol{\Psi} \in \mathcal{P}} \|\boldsymbol{\Psi}\|_0 \quad \mathrm{s.t.} \quad \inf_{x \in (0, \eta]} \pmb{\hbar}^T \mho(x) \boldsymbol{\Psi} + \theta \ln(1-x) \ge 0.
\]
A common relaxation technique is to replace the $\ell_0$-norm into a $\ell_1$-norm.
However, this does not work in our problem, as $\boldsymbol{\Psi}$ is a probability distribution so that we must have $\|\boldsymbol{\Psi}\|_1 = 1$.
This way, the relaxation gives a feasible $\boldsymbol{\Psi}$ with arbitrary sparsity.

A variation of the relaxation is the iterative reweighted $\ell_1$-norm heuristic.
The standard approach is to approximate $\mathbbm{1}_{x > 0}$ by $\log(1+x/\alpha)$ for some parameter $\alpha > 0$.
In our case, all $\Psi_d$ are small, so we want an approximation of $\mathbbm{1}_{\Psi_d > 0}$ that grows fast for small $\Psi_d$.
Therefore, we scale the logarithm to get the approximation $\mathbbm{1}_{\Psi_d > 0} \approx \delta^{-1} \ln ( 1 + \Psi_d / \alpha )$
for some $\delta, \alpha> 0$.
On the other hand, we want the approximation equals $1$ when $\Psi_d = 1$ so that the approximation is more accurate.
That is, we desire $\delta^{-1} \ln ( 1 + 1 / \alpha ) = 1$, which gives $\alpha = (e^\delta - 1)^{-1}$.
To conclude, we choose the approximation 
\begin{IEEEeqnarray*}{rCl}
    \| \boldsymbol{\Psi} \|_0 & \approx & \sum_{d = 1}^D \frac{1}{\delta} \ln ( 1 + (e^\delta - 1) \Psi_d) \yesnumber \label{eq:l1log}\\
    & \approx & \frac{1}{\delta} \sum_{d = 1}^D \ln(1 + (e^\delta - 1) \Psi_d^{(k)}) + \frac{1}{\delta} \sum_{d = 1}^D \frac{\Psi_d - \Psi_d^{(k)}}{\frac{1}{e^\delta - 1} + \Psi_d^{(k)}},
\end{IEEEeqnarray*}
where the last approximation is the linearization at the point $\boldsymbol{\Psi}^{(k)} = (\Psi_1^{(k)}, \Psi_2^{(k)}, \ldots, \Psi_D^{(k)})$.
The value $k$ denotes the $k$-th iteration that will be discussed later.

As the linearized function majorizes the first approximation \eqref{eq:l1log}, we can optimize the linearized form instead of \eqref{eq:l1log} iteratively as an majorization-minimization algorithm.
By ignoring the constant terms, we obtain the optimization program
\begin{equation} \label{eq:l1opt}
    \min_{\boldsymbol{\Psi} \in \mathcal{P}} \mathbf{w}^T \boldsymbol{\Psi} \quad \mathrm{s.t.} \quad \inf_{x \in (0, \eta]} \pmb{\hbar}^T \mho(x) \boldsymbol{\Psi} + \theta \ln(1-x) \ge 0,
\end{equation}
where $w_d = (\frac{1}{e^\delta - 1} + \Psi_d^{(k)})^{-1}$ for all $d \in \llbracket D \rrbracket$.

In the $k$-th iteration, we solve the above optimization problem to obtain a $\boldsymbol{\Psi}$.
Then, we update the weights $w_d$ by $(\frac{1}{e^\delta - 1} + \Psi_d)^{-1} / \delta$.
The scaling factor $1/\delta$ is not mandatory.
The purpose is to reduce the size of $w_d$, which can be large as $\frac{1}{e^\delta - 1}$ and $\Psi_d$ are small.
We break the iteration if the weight $\mathbf{w}$ converges, say, the $\ell_1$-norm of the change of $\mathbf{w}$ is smaller than some $\epsilon_1 > 0$.
Otherwise, we set $\boldsymbol{\Psi}^{(k+1)}$ to be $\boldsymbol{\Psi}$.
On the other hand, we fix a maximum number of iterations $k_\text{max}$ as the heuristic does not guarantee the convergence of $\mathbf{w}$.

After all iterations are done, we work on the latest $\boldsymbol{\Psi}$.
We trim the masses smaller than a threshold $\epsilon_2 > 0$ to $0$, and then normalize the non-zero masses.
This way, we obtain a sparse $\boldsymbol{\Psi}$.
The whole procedure is described in \Cref{Alg:l1}.

\begin{figure}
\centering
\removelatexerror
\begin{algorithm}[H]
\small
  \SetAlgoLined
  \SetKwInOut{Input}{Input}
  \SetKwInOut{Output}{Output}
  \ResetInOut{Output}
  \Input{Target rate $\theta$, initial guess $\boldsymbol{\Psi}^{(0)}$}
  \Output{A sparse degree distribution $\boldsymbol{\Psi}$}

  $\mathbf{w} \gets \mathbf{1}_{D}$, $k \gets 1$, $\boldsymbol{\Psi}^{(1)} \gets \boldsymbol{\Psi}^{(0)}$\;
  \While{$k \le k_\text{max}$}{
    Solve \eqref{eq:l1opt} to obtain $\boldsymbol{\Psi}$\;
    \lIf{Failed to solve \eqref{eq:l1opt}}{
      Break the loop
    }
    $\mathbf{w}'_d \gets (\frac{1}{e^\delta - 1} + \Psi_d)^{-1} / \delta$ for all $d \in \llbracket D \rrbracket$\;
    \lIf{$\|\mathbf{w}' - \mathbf{w}\|_1 < \epsilon_1 $}{
      Break the loop
    }
    $\mathbf{w} \gets \mathbf{w}'$, $\boldsymbol{\Psi}^{(k+1)} \gets \boldsymbol{\Psi}$, $k \gets k+1$\;
  }
  $\boldsymbol{\Psi} \gets \boldsymbol{\Psi}^{(k)}$\;
  Set $\Psi_d = 0$ if $\Psi_d < \epsilon_2$ for all $d \in \llbracket D \rrbracket$\;
  $\boldsymbol{\Psi} \gets \boldsymbol{\Psi}/ \mathbf{1}_D^T \boldsymbol{\Psi}$ \;
  \KwRet{$\boldsymbol{\Psi}$}\;
  
  \caption{Iterative Reweighted $\ell_1$-Norm Heuristic}\label{Alg:l1}
\end{algorithm}
\vspace{-1.5em}
\end{figure}

\subsection{Exact Solver}\label{Sec:exact:solver}

Suppose we want to restrict the size of the support of the degree distribution $\boldsymbol{\Psi}$ to be at most $s$.
Define the probability simplex
    $\mathcal{P}' = \left\{ 
    \boldsymbol{\Psi} \in \mathcal{P} \colon \|\boldsymbol{\Psi}\|_0\le s
    \right\}.$
Let $\mathcal{X}$ be the discretized set of the interval $(0, \eta]$.
This way, we can introduce a sparsity constraint to \eqref{Eq:optimization:nominal} and rewrite it as
\begin{equation}
  \max_{\theta \in \mathbb{R}} \,\,\, \theta \quad \mathrm{s.t.} \quad \max_{\boldsymbol{\Psi} \in \mathcal{P}'} \min_{x \in \mathcal{X}} (\pmb{\hbar}^T \mho(x) \boldsymbol{\Psi} + \theta \ln(1-x)) \ge 0.\label{Eq:optimization:discrete}
\end{equation}
This formulation is a mixed-integer linear programming problem, which is NP-hard in general.
We apply bisection search to find the largest possible $\theta$ such that the constraint is satisfied.
This way, 
the remaining question is to find a tractable algorithm to check whether a given $\theta$ is feasible, which suffices to compute the optimal value of $T(\theta):=\max_{\boldsymbol{\Psi} \in \mathcal{P}'} \min_{x \in \mathcal{X}} (\pmb{\hbar}^T \mho(x) \boldsymbol{\Psi} + \theta \ln(1-x)).$

This formulation is a sparse optimization with respect to a piecewise linear function, which is again NP-hard in general.

The main difficulty for solving 
$T(\theta)$
is that the constraint set $\mathcal{P}'$ involves an $\ell_0$-norm constraint $\|\Psi\|_0\le s$.
To tackle this issue, we introduce a binary vector $\mathbf{z} \in \{0,1\}^D$ such that $z_d = \mathbbm{1}_{\Psi_d \neq 0}$ for all $d \in \llbracket D \rrbracket$.
Define the set
    $\mathcal{Z} := \{ \mathbf{z} \in \{0,1\}^D \colon \| \mathbf{z} \|_0 \le s \}$.
Hence, we linearize this $\ell_0$-norm constraint using the binary vector $\mathbf{z}$, such that $T(\theta)$ is reformulated as the following mixed-integer program:
\begin{multline*}
    \max_{\boldsymbol{\Psi} \in \mathcal{P}, \mathbf{z} \in \mathcal{Z}, \tau \in \mathbb{R}} \tau\\
    \mathrm{s.t.} \,\,\, \boldsymbol{\Psi} \le \mathbf{z}, \quad \tau \le \pmb{\hbar}^T \mho(x) \boldsymbol{\Psi} + \theta \ln(1-x), \forall x \in \mathcal{X}.
\end{multline*}
There can be many constraints as $|\mathcal{X}|$ can be huge, thus solving this problem directly can be inefficient.

Next, we apply the Benders decomposition to separate the binary variables $\mathbf{z}$.
Write the above mixed-integer problem as
\begin{equation}\label{Eq:max:fz}
    \max_{\mathbf{z} \in \mathcal{Z}} \, f(\mathbf{z}),
\end{equation}
where $f(\mathbf{z})$ is the optimal value of the linear program
\begin{multline*}
    \max_{\boldsymbol{\Psi} \in \mathcal{P}, \tau \in \mathbb{R}} \tau\\
    \mathrm{s.t.} \, \boldsymbol{\Psi} \le \mathbf{z}, \quad \tau \le \pmb{\hbar}^T \mho(x) \boldsymbol{\Psi} + \theta \ln(1-x), \forall x \in \mathcal{X}.
\end{multline*}
By strong duality, we can rewrite the above problem as 
\begin{multline} \label{Eq:fz:dual}
    \min_{\boldsymbol{\alpha} \ge \mathbf{0}_{|\mathcal{X}|}, \boldsymbol{\beta} \ge \mathbf{0}_D, \mu \in \mathbb{R}} \, \sum_{x \in \mathcal{X}} \alpha_x\theta\ln(1-x) + \boldsymbol{\beta}^T \mathbf{z} + \mu\\
    \mathrm{s.t.} \, \sum_{x \in \mathcal{X}} \alpha_x = 1, \quad \sum_{x \in \mathcal{X}} \alpha_x \pmb{\hbar}^T \mho(x) \le \mu \mathbf{1}_D^T + \boldsymbol{\beta}.
\end{multline}
When $\mathbf{z}$ is fixed, we can evaluate $f(\mathbf{z})$, i.e., solving the above dual problem, efficiently.
The remaining problem is to solve \eqref{Eq:max:fz}.

One can see that the objective $f(z)$ defined in \eqref{Eq:fz:dual} is concave in $z$, since it can be viewed as the minimization of (infinitely many) linear functions in $z$.
This motivates us to provide an outer approximation algorithm to solve \eqref{Eq:max:fz} within a desired global optimality gap.
Its high-level idea is to iteratively optimize a piecewise linear approximation of the objective $f$ and refining this approximation based on its subgradient, until a certain approximation error is achieved.
We summarize its detailed algorithm in \Cref{Alg:outer:alg}.
This algorithm has been originally proposed in \cite{kelley1960cutting} for continuous decision variables, and later was extended for generic optimization with binary variables~\cite{duran1986outer}.
The authors therein also pointed out that this algorithm is guaranteed to converge in finite, yet exponential in worst-case, number of iterations.
Practical evaluation suggests that the problem usually converges in a few iterations.
Therefore, the remaining mixed-integer program, i.e., the one inside \Cref{Alg:outer:alg}, involves a few constraints only.

It is worth mentioning that \Cref{Alg:outer:alg} requires the oracle of a subgradient of $f$, which can be obtained by Danskin's theorem as follows:
For a fixed $\mathbf{z}$, we first solve Problem~\eqref{Eq:fz:dual} to obtain an optimal solution $(\boldsymbol{\alpha}(\mathbf{z}), \boldsymbol{\beta}(\mathbf{z}), \mu(\mathbf{z}))$ and then construct a subgradient of $f(\mathbf{z})$ as
$g(\mathbf{z}) := \boldsymbol{\beta}(\mathbf{z})$.

\begin{figure}
\centering
\removelatexerror
\begin{algorithm}[H]
\small
  \SetAlgoLined
  \SetKwInOut{Input}{Input}
  \SetKwInOut{Output}{Output}
  \ResetInOut{Output}
  \Input{Initial guess $z^{(1)}$, accuracy level $\epsilon$}
  \Output{$\epsilon$-optimal sparsity vector $\mathbf{z}$}

  $t\leftarrow 1$, $\textit{Error}\leftarrow \infty$\;
  Compute $f(\mathbf{z}^{(1)})$ and its subgradient $g(\mathbf{z})^{(1)}$\;
  \While{$|\text{Error}| > \epsilon$}{
    Compute the optimal $\mathbf{z}^{(t+1)}$ and $\Upsilon^{(t+1)}$ of 
    \begin{multline*}
        \max_{\mathbf{z}^{(t+1)}\in\mathcal{Z}, \Upsilon^{(t+1)} \in \mathbb{R}} \Upsilon^{(t+1)}\\
        \mathrm{s.t.} \, \Upsilon^{(t+1)} \le f(\mathbf{z}^{(s)}) + 
        \langle \mathbf{z}-\mathbf{z}^{(s)},
        g(\mathbf{z}^{(s)}) \rangle, \forall s\in \llbracket t \rrbracket;
    \end{multline*}
        
    Compute $f(\mathbf{z}^{(t+1)})$ and its subgradient $g(\mathbf{z}^{(t+1)})$\;
    Update $\textit{Error}\leftarrow \Upsilon^{(t+1)} - f(\mathbf{z}^{(t+1)})$\;
    Update $t\leftarrow t+1$\;
  }
  \KwRet{$\mathbf{z}^{(t)}$}\;
  
  \caption{Optimal Sparsity Vector via Problem~\eqref{Eq:max:fz}}\label{Alg:outer:alg}
\end{algorithm}
\vspace{-1.5em}
\end{figure}

\section{Numerical Evaluations}

In this section, we examine the performance of several approaches proposed in Section~\ref{Sec:sparse:opt} for $\eta=0.98$ and $0.99$.
We take $|\mathcal{X}|=200$ when implementing the exact solvers.
For the other methods, we discretize $\mathcal{X}$ with a step size of $0.001$, using the SeDuMi solver on MATLAB CVX on an Intel Core i5-6500 3.20GHz, 32GB RAM, Windows 10 machine.
In the heuristics, we use a threshold $10^{-7}$ to trim down the small entries.
For the $\ell_1$-norm heuristic, we use $\delta = 10$, $k_\text{max} = 10$, and $\epsilon_1 = 10^{-3}$.

We quantify the performance of a given degree distribution $\boldsymbol{\Psi}$ using the relative error between its achievable rate and the rate obtained from optimal degree optimization without any sparsity constraints, called the \emph{rate drop}.
A smaller rate drop corresponds to the superior performance of the obtained degree distribution. 
To illustrate the trend for comparison, we consider a binomial distribution $B(M, 0.8)$ as the rank distribution with $M=8$.
We report the rate drop, computational time, and number of non-zero entries in degree distribution for each approach in Table~\ref{Tab:achievable:rate:summary}.
We have the following observations regarding our proposed approaches:
\begin{enumerate}
    \item 
Heuristics in Sections~\ref{Sec:direct} to \ref{Sec:ell:1} have fast computational speed, but their corresponding performances may not reach the best levels. Also, they cannot be used to optimize for a specific support size.
    \item
Direct trimming approach results in a huge support size, thus it is not a desired approach but for comparison purpose only.
    \item
Complementary slackness approach has the fastest running time, thus it is a candidate for when we need to produce new degree distribution frequently due to the change of channel condition. Also, it is suitable to be used in devices with limited computational power.
\item
    $\ell_1$-norm heuristic has a smaller rate drop than the complementary slackness approach. However, the running time is much longer. It is suitable when the degree distribution can be reused for a sufficiently long time frame, e.g., stable channel condition.
    \item
The exact solver yields the sparse distribution that excels in the case of $\eta=0.98$ and approaches the best performance for $\eta=0.99$, albeit at the expense of the highest computational cost among all methods.
Additional numerical studies are provided in 
\ifnum\paperversion=1
the Appendix of the full version \cite{full_sparse}
\fi
\ifnum\paperversion=2
Appendix~\ref{Sec:num:add}
\fi
to demonstrate the efficiency and high-quality solution using the exact solver proposed in Section~\ref{Sec:exact:solver}.
\end{enumerate}
\begin{table} 
\vskip 1em
\caption{
Achievable rates and computational time for different approaches} %with various $\eta$ values.}
\label{Tab:achievable:rate:summary}
    \centering
        \begin{tabular}{c|ccc}
        \hline
        \multirow{2}{*}{Method} & \multicolumn{3}{c}{$\eta=0.98$} \\[3pt] 
        \cline{2-4} %\\[0.1pt] 
        & Rate Drop & Time & Support Size \\[3pt]
        \hline
       $\begin{array}{c}
            \mbox{Direct}\\
            \mbox{Trimming}
       \end{array}$  & 3.15e-6 & 6.63 & 154\\
        \hline
       $\begin{array}{c}
            \mbox{Complementary}\\
            \mbox{Slackness}
       \end{array}$  & 6.32e-7 & 2.54 & 14 \\
        \hline
       $\begin{array}{c}
            \mbox{$\ell_1$-Norm}\\
            \mbox{Heuristic}
       \end{array}$  & 7.25e-5 & 66.27 & 11\\
       \hline
       $\begin{array}{c}
            \mbox{Exact}\\
            \mbox{Solver}
       \end{array}$ &5.42e-7 &460.17 & 12\\\hline
        $\begin{array}{c}
            \mbox{Optimal}\\
            \mbox{(Non-sparse)}
       \end{array}$  & - & 6.45 & 235  \\[5pt]
       \hline
    \end{tabular}\\\vspace{0.3em}
     \begin{tabular}{c|ccc}
        \hline
        \multirow{2}{*}{Method} & \multicolumn{3}{c}{$\eta=0.99$} \\[3pt] 
        \cline{2-4} %\\[0.1pt] 
        & Rate Drop & Time & Support Size \\[3pt]
        \hline
       $\begin{array}{c}
            \mbox{Direct}\\
            \mbox{Trimming}
       \end{array}$  & 2.55e-5 & 13.28 & 299\\
        \hline
       $\begin{array}{c}
            \mbox{Complementary}\\
            \mbox{Slackness}
       \end{array}$  & 3.30e-7 & 7.81 & 16\\
        \hline
       $\begin{array}{c}
            \mbox{$\ell_1$-Norm}\\
            \mbox{Heuristic}
       \end{array}$  & 4.86e-5 & 131.02 & 13\\
       \hline
       $\begin{array}{c}
            \mbox{Exact}\\
            \mbox{Solver}
       \end{array}$ & 6.41e-7&368.52 & 12\\\hline
        $\begin{array}{c}
            \mbox{Optimal}\\
            \mbox{(Non-sparse)}
       \end{array}$  & - & 13.16 & 489 \\[5pt]
       \hline
    \end{tabular}
    \vspace{-2em}
    \label{tab:mytable}
\end{table}

\section{Concluding Remarks}
In this paper, we investigate the sparse optimization of degree distribution for BATS codes and propose various heuristics and an exact solver.
We find heuristic approaches typically have faster computational speed, while they neither have satisfactory performance nor allow tuning the support size.
Although the exact solver incurs much larger computational cost, it returns degree distribution with satisfactory performance and controlled support size.
These approaches give a trade-off between computational time and achievable rate, thus giving us the flexibility to adapt to various scenarios.

\ifnum\paperversion=1
\clearpage
\balance
\bibliographystyle{IEEEtran}
\bibliography{ref}

\end{document}
\fi

\appendices

\section{Proofs of Technical Results} \label{Sec:proof:converge}
\begin{IEEEproof}[Proof of Proposition~\ref{Proposition:converge}]
The key is to re-write Problem~\eqref{Eq:optimization:nominal} as a semi-infinite programming:
\begin{equation}
    \smashoperator{\max_{\,\,\,\boldsymbol{\Psi} \in \mathcal{P}', \theta \in \mathbb{R}}} \,\,\theta \,\,\, \mathrm{s.t.} \,\,\, (\pmb{\hbar}^T \mho(x) \boldsymbol{\Psi} + \theta \ln(1-x)) \ge 0, \forall x \in [0, \eta].
    \label{Eq:optimization:nominal:revised}
\end{equation}
It is worth mentioning that we replace the interval $(0,\eta]$ with $[0,\eta]$, since $x=0$ also satisfies $G(\theta,\boldsymbol{\Psi}; x) = (\pmb{\hbar}^T \mho(x) \boldsymbol{\Psi} + \theta \ln(1-x)) \ge 0$.
It can be verified that $G$ is continuous in $\mathbb{R}\times \mathcal{P}\times [0,\eta]$.
The result follows directly by applying \cite[Lemma~6.1]{shapiro2009semi}.
\end{IEEEproof}
\begin{IEEEproof}[Proof of Theorem~\ref{Theorem:converge:rate}]
As has been shown that Problem~\eqref{Eq:optimization:nominal} is a semi-infinite programming, one can apply \cite[Theorem~6.1]{shapiro2009semi} to derive the convergence rate of discretization.
\end{IEEEproof}

\section{Additional Numerical Study}\label{Sec:num:add}
\begin{figure}
    \centering
    \includegraphics[width=0.38\textwidth]{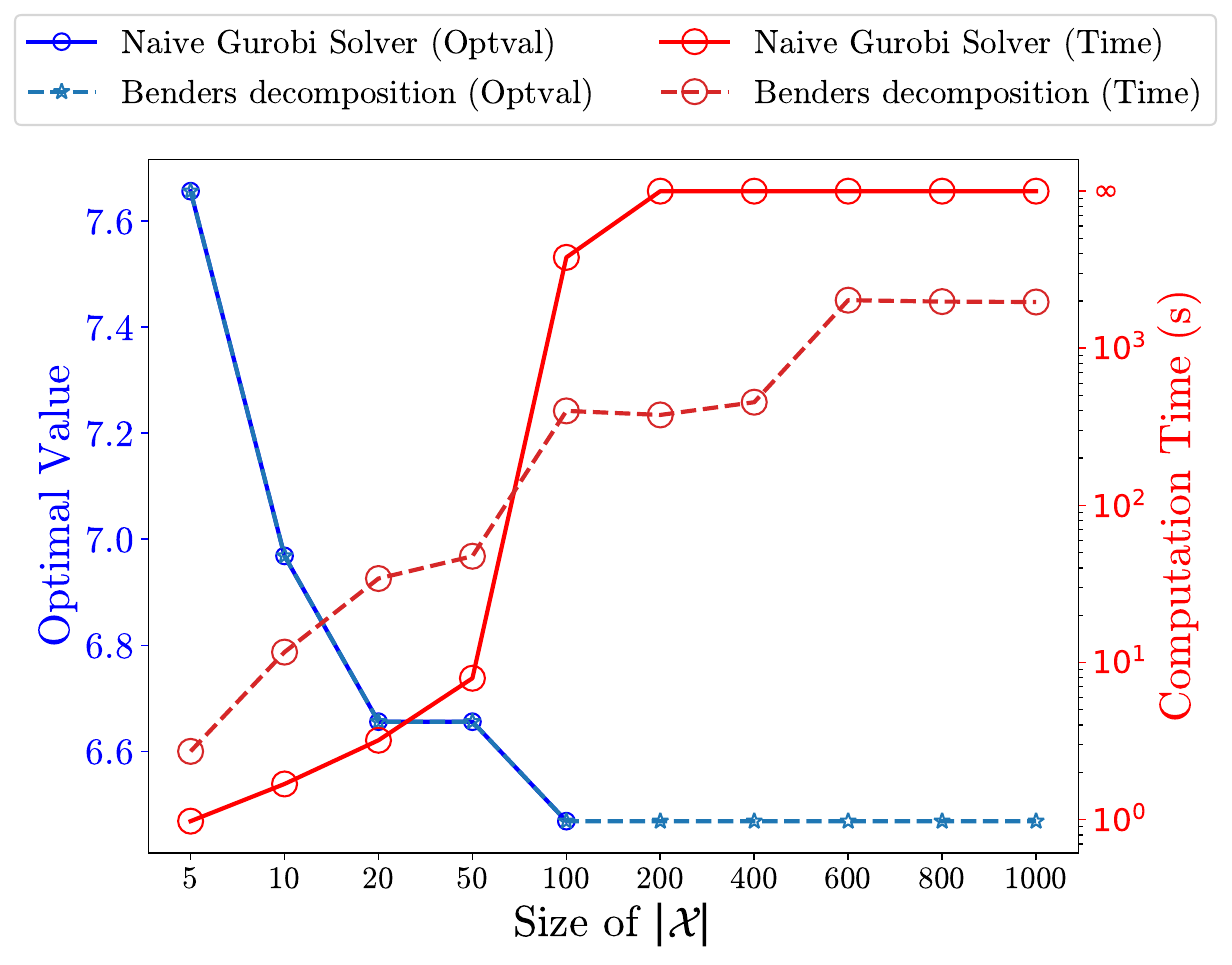}
    \caption{Comparison of optimal value (in left $y$-axis) and computational time (in right $y$-axis) with various choices of $|\mathcal{X}|$ for exact degree optimization solver.
    The naive Gurobi solver successfully solves only the first $5$ instances within $10^4$ seconds, whereas our customized algorithm is quite scalable for large problem size.
    }
    \label{fig:compute:optval}
\end{figure}

We first evaluate the computational efficiency by investigating the computational time of the exact solver.
This comparison is conducted between our proposed algorithm in Section~\ref{Sec:exact:solver}, and a naive Gurobi solver directly applied to solve $T(\theta)$.
The termination criterion for the exact solver is set to either the time exceeds $10^4$ seconds or find a solution $\hat{\theta}$ such that the relative error, expressed as $\frac{|\hat{\theta} - \theta_*|}{|\theta^{u} - \theta^l|}$, falls below $1\%$.
Here,  $\theta_*$ denotes the optimal solution to \eqref{Eq:optimization:discrete}, and the initial interval for bisection search is defined as $[\theta^l, \theta^u]=[0,8]$.
Numerical results with varying sizes of $|\mathcal{X}|$ are presented in Fig.~\ref{fig:compute:optval}, which indicates that with increasing $|\mathcal{X}|$, the optimal value of \eqref{Eq:optimization:discrete} tends to converge~(i.e., when $|\mathcal{X}|\ge100$), aligning with our theoretical findings in Proposition~\ref{Proposition:converge} and Theorem~\ref{Theorem:converge:rate}.
Notably, as $|\mathcal{X}|$ increases, the computational time of the naive Gurobi solver experiences a substantial increase, whereas our customized algorithm exhibits remarkable scalability.

\begin{figure}
    \centering
    \includegraphics[width=0.38\textwidth]{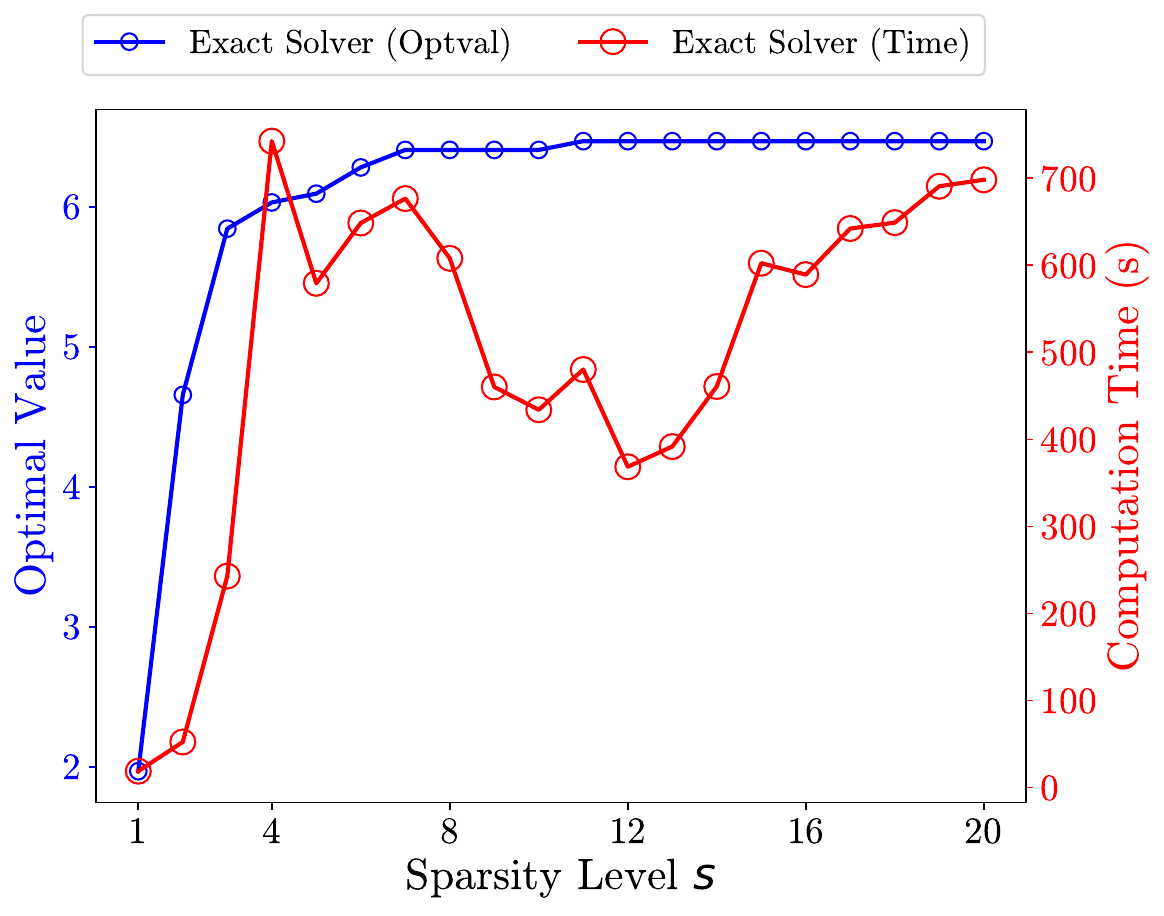}
    \caption{Comparison of optimal value (in left $y$-axis) and computational time (in right $y$-axis) with various choices of sparsity level $s$ for exact degree optimization solver.
    }
    \label{fig:compute:sparse}
\end{figure}

Next, we perform an ablation study to examine the impact of sparsity level $s$ on the performance of the exact solver in Fig.~\ref{fig:compute:sparse}.
The plot indicates that as the sparsity level $s$ increases, the achievable rate for the exact solver tends to converge, especially when $s\ge11$.
This justifies the fact that setting a sparse degree distribution leads to a close-to-optimal rate.

\balance
\bibliographystyle{IEEEtran}
\bibliography{ref}

\begin{thebibliography}{10}
\providecommand{\url}[1]{#1}
\csname url@samestyle\endcsname
\providecommand{\newblock}{\relax}
\providecommand{\bibinfo}[2]{#2}
\providecommand{\BIBentrySTDinterwordspacing}{\spaceskip=0pt\relax}
\providecommand{\BIBentryALTinterwordstretchfactor}{4}
\providecommand{\BIBentryALTinterwordspacing}{\spaceskip=\fontdimen2\font plus
\BIBentryALTinterwordstretchfactor\fontdimen3\font minus
  \fontdimen4\font\relax}
\providecommand{\BIBforeignlanguage}[2]{{%
\expandafter\ifx\csname l@#1\endcsname\relax
\typeout{** WARNING: IEEEtran.bst: No hyphenation pattern has been}%
\typeout{** loaded for the language `#1'. Using the pattern for}%
\typeout{** the default language instead.}%
\else
\language=\csname l@#1\endcsname
\fi
#2}}
\providecommand{\BIBdecl}{\relax}
\BIBdecl

\bibitem{linear}
S.-Y.~R. Li, R.~W. Yeung, and N.~Cai, ``Linear network coding,'' \emph{IEEE
  Trans. Inf. Theory}, vol.~49, no.~2, pp. 371--381, Feb. 2003.

\bibitem{flow}
R.~Ahlswede, N.~Cai, S.-Y.~R. Li, and R.~W. Yeung, ``Network information
  flow,'' \emph{IEEE Trans. Inf. Theory}, vol.~46, no.~4, pp. 1204--1216, Jul.
  2000.

\bibitem{random2}
T.~Ho, M.~M{\'e}dard, R.~Koetter, D.~R. Karger, M.~Effros, J.~Shi, and
  B.~Leong, ``A random linear network coding approach to multicast,''
  \emph{IEEE Trans. Inf. Theory}, vol.~52, no.~10, pp. 4413--4430, Oct. 2006.

\bibitem{yang14bats}
S.~Yang and R.~W. Yeung, ``Batched sparse codes,'' \emph{IEEE Trans. Inf.
  Theory}, vol.~60, no.~9, pp. 5322--5346, Sep. 2014.

\bibitem{bats_book}
------, \emph{{{BATS} Codes: Theory and Practice}}, ser. Synthesis Lectures on
  Communication Networks.\hskip 1em plus 0.5em minus 0.4em\relax Morgan \&
  Claypool Publishers, 2017.

\bibitem{lubyLT}
M.~Luby, ``{LT} codes,'' in \emph{Proc. FOCS '02}, Nov. 2002, pp. 271--282.

\bibitem{quasi}
X.~Xu, Y.~L. Guan, Y.~Zeng, and C.-C. Chui, ``Quasi-universal {BATS} code,''
  \emph{IEEE Trans. Veh. Technol.}, vol.~66, no.~4, pp. 3497--3501, 2017.

\bibitem{expanding}
X.~{Xu}, Y.~{Zeng}, Y.~L. {Guan}, and L.~{Yuan}, ``Expanding-window {BATS} code
  for scalable video multicasting over erasure networks,'' \emph{IEEE Trans.
  Multim.}, vol.~20, no.~2, pp. 271--281, Feb. 2018.

\bibitem{sliding}
J.~{Yang}, Z.~{Shi}, C.~{Wang}, and J.~{Ji}, ``Design of optimized
  sliding-window {BATS} codes,'' \emph{IEEE Commun. Lett.}, vol.~23, no.~3, pp.
  410--413, Mar. 2019.

\bibitem{unequal}
X.~Xu, Y.~Zeng, Y.~L. Guan, and L.~Yuan, ``{BATS} code with unequal error
  protection,'' in \emph{Proc. ICCS'16}, Dec. 2016.

\bibitem{yue2018distributed}
J.~Yue, M.~Xiao, and Z.~Pang, ``Distributed fog computing based on batched
  sparse codes for industrial control,'' \emph{IEEE Trans. Ind. Informat.},
  vol.~14, no.~10, pp. 4683--4691, 2018.

\bibitem{shokRaptor}
A.~Shokrollahi, ``{R}aptor codes,'' \emph{IEEE Trans. Inf. Theory}, vol.~52,
  no.~6, pp. 2551--2567, Jun. 2006.

\bibitem{pktsize}
H.~H.~F. Yin, H.~W.~H. Wong, M.~Tahernia, and J.~Qing, ``Packet size
  optimization for batched network coding,'' in \emph{Proc. ISIT '22}, Jun.
  2022, pp. 1584--1589.

\bibitem{yang22pro}
S.~Yang and R.~W. Yeung, ``Network communication protocol design from the
  perspective of batched network coding,'' \emph{IEEE Commun. Mag.}, vol.~60,
  no.~1, pp. 89--93, Jan. 2022.

\bibitem{yin20pro}
H.~H.~F. Yin, R.~W. Yeung, and S.~Yang, ``A protocol design paradigm for
  batched sparse codes,'' \emph{Entropy}, vol.~22, no.~7, Jul. 2020, {A}rt. no.
  790.

\bibitem{variable}
Q.~Zhou, S.~Yang, H.~H.~F. Yin, and B.~Tang, ``On {BATS} codes with variable
  batch sizes,'' \emph{IEEE Commun. Lett.}, vol.~21, no.~9, pp. 1917--1920,
  Sep. 2017.

\bibitem{fun}
H.~Zhang, K.~Sun, Q.~Huang, Y.~Wen, and D.~Wu, ``{FUN} coding: Design and
  analysis,'' \emph{IEEE/ACM Trans. Netw.}, vol.~24, no.~6, pp. 3340--3353,
  Dec. 2016.

\bibitem{delay}
H.~H.~F. Yin, K.~H. Ng, X.~Wang, and Q.~Cao, ``On the minimum delay of block
  interleaver for batched network codes,'' in \emph{Proc. ISIT'19}, Jul. 2019,
  pp. 1957--1961.

\bibitem{buffer}
H.~H.~F. Yin, K.~H. Ng, X.~Wang, Q.~Cao, and L.~K.~L. Ng, ``On the memory
  requirements of block interleaver for batched network codes,'' in \emph{Proc.
  ISIT'20}, Jun. 2020, pp. 1658--1663.

\bibitem{zhou17b}
Z.~Zhou, C.~Li, S.~Yang, and X.~Guang, ``Practical inner codes for {BATS} codes
  in multi-hop wireless networks,'' \emph{IEEE Trans. Veh. Technol.}, vol.~68,
  no.~3, pp. 2751--2762, Mar. 2019.

\bibitem{bats_schedule}
Z.~Zhou, J.~Kang, and L.~Zhou, ``Joint {BATS} code and periodic scheduling in
  multihop wireless networks,'' \emph{IEEE Access}, vol.~8, pp.
  29\,690--29\,701, Feb. 2020.

\bibitem{yang14a}
S.~Yang, R.~W. Yeung, J.~H.~F. Cheung, and H.~H.~F. Yin, ``{BATS}: Network
  coding in action,'' in \emph{Proc. ALLERTON '14}, Sep. 2014, pp. 1204--1211.

\bibitem{adaptive}
H.~H.~F. Yin, S.~Yang, Q.~Zhou, and L.~M.~L. Yung, ``Adaptive recoding for
  {BATS} codes,'' in \emph{Proc. ISIT '16}, Jul. 2016, pp. 2349--2353.

\bibitem{scheduling}
B.~Tang, S.~Yang, B.~Ye, S.~Guo, and S.~Lu, ``Near-optimal one-sided scheduling
  for coded segmented network coding,'' \emph{IEEE Trans. Comput.}, vol.~65,
  no.~3, pp. 929--939, Mar. 2016.

\bibitem{ge_adaptive}
X.~Xu, Y.~L. Guan, and Y.~Zeng, ``Batched network coding with adaptive recoding
  for multi-hop erasure channels with memory,'' \emph{IEEE Trans. Commun.},
  vol.~66, no.~3, pp. 1042--1052, Mar. 2018.

\bibitem{uni}
H.~H.~F. Yin, B.~Tang, K.~H. Ng, S.~Yang, X.~Wang, and Q.~Zhou, ``A unified
  adaptive recoding framework for batched network coding,'' \emph{IEEE JSAIT},
  vol.~2, no.~4, pp. 1150--1164, Dec. 2021.

\bibitem{bar}
H.~H.~F. Yin, S.~Yang, Q.~Zhou, L.~M.~L. Yung, and K.~H. Ng, ``{BAR}: Blockwise
  adaptive recoding for batched network coding,'' \emph{Entropy}, vol.~25,
  no.~7, Jul. 2023, {A}rt. no. 1054.

\bibitem{Raptormono}
A.~Shokrollahi and M.~Luby, \emph{Raptor Codes}, ser. Foundations and Trends in
  Communications and Information Theory.\hskip 1em plus 0.5em minus 0.4em\relax
  now, 2011, vol.~6.

\bibitem{yang11x2}
S.~Yang, S.-W. Ho, J.~Meng, and E.-H. Yang, ``Capacity analysis of linear
  operator channels over finite fields,'' \emph{IEEE Trans. Inf. Theory},
  vol.~60, no.~8, pp. 4880--4901, Aug. 2014.

\bibitem{gadouleau10packing}
M.~Gadouleau and Z.~Yan, ``Packing and covering properties of subspace codes
  for error control in random linear network coding,'' \emph{IEEE Trans. Inf.
  Theory}, vol.~56, no.~5, pp. 2097--2108, May 2010.

\bibitem{tree}
S.~Yang and Q.~Zhou, ``Tree analysis of {BATS} codes,'' \emph{IEEE Commun.
  Lett.}, vol.~20, no.~1, pp. 37--40, Jan. 2016.

\bibitem{kelley1960cutting}
J.~E. Kelley, Jr, ``The cutting-plane method for solving convex programs,''
  \emph{Journal of the Society for Industrial and Applied Mathematics}, vol.~8,
  no.~4, pp. 703--712, 1960.

\bibitem{duran1986outer}
M.~A. Duran and I.~E. Grossmann, ``An outer-approximation algorithm for a class
  of mixed-integer nonlinear programs,'' \emph{Math. Program.}, vol.~36, pp.
  307--339, 1986.

\bibitem{shapiro2009semi}
A.~Shapiro, ``Semi-infinite programming, duality, discretization and optimality
  conditions,'' \emph{Optimization}, vol.~58, no.~2, pp. 133--161, 2009.

\end{thebibliography}

\end{document}